\documentclass[12pt]{article}

\usepackage{amsmath}
\usepackage{amssymb}

\usepackage{graphicx}

\usepackage{cite}

\usepackage{color} 

\usepackage[labelfont=bf,labelsep=period,justification=raggedright]{caption}

\makeatletter
\renewcommand{\@biblabel}[1]{\quad#1.}
\makeatother

\date{}

\begin{document}

% Title must be 150 characters or less
\begin{flushleft}
{\Large
\textbf{Interventions in Networks}
}
% Insert Author names, affiliations and corresponding author email.
\\
Steven K. Thompson$^{1}$, 
\\
\bf{1} 
 Department of Statistics and Actuarial Science, Simon Fraser
 University, Burnaby, BC, Canada
\\
$\ast$ E-mail: thompson@sfu.ca
\end{flushleft}

% Please keep the abstract between 250 and 300 words
\section*{Abstract}

Interventions are made in networks to change the network or its values
in a desired way.  The intervention strategies evaluated in the study
described here use network sampling designs to find units to which
interventions are applied.  An intervention applied to a network node
or link can change a value associated with that unit.  Over time the
effect of the intervention can have an effect on the population that
goes beyond the sample units to which it is directly applied.  This
paper describes the methods used for this study.  These include a
variety of link-tracing sampling designs in networks, a number of
types of interventions, and a temporal spatial network model in which
the intervention strategies are evaluated.  An intervention strategy
is associated with an agent and different intervention strategies
interact and adapt to each other over time.  Some preliminary results
are summarized regarding potential intervention strategies to help
alleviate the HIV epidemic.

% Please keep the Author Summary between 150 and 200 words
% Use first person. PLoS ONE authors please skip this step. 
% Author Summary not valid for PLoS ONE submissions.   
\section*{Author Summary}

\section*{Introduction}

The purpose of the project described here is to create a sampling
design and simulation system for evaluating the effectiveness of
potential strategies for interventions in networks.  The motivating
problem for this work has been the effort to understand and reduce the
spread of HIV.  An effective intervention is one that helps
in reducing the epidemic at least locally in an area of focus.

The approach taken uses a network or spatial sampling design to find
units to which to make interventions.  These designs in many cases
involve tracing of links from sample units to add new units to the
sample.  The population network itself is the result of choices of
individuals, and consensus of pairs of individuals, in forming and
dissolving links between them.  The tendencies of individuals in this
process are viewed also as sampling designs.

A typical design in this study might use spatial sampling to find
initial candidate nodes for selection.  Subsequent selections can use
link tracing to find additional candidate nodes.  The sampling process
includes a selection process for adding units to the sample and an
attrition process by which nodes are removed from the sample.

An intervention strategy is the combination of a sampling design and
the intervention made to its sample units.  We think of a design or
strategy as having an agent behind it.  Sexually transmissible
infections are often countered by public health agencies with seek and
treat designs which follow sexual links from infected persons and test
and treat their recent partners.  A person choosing not to take on
additional partners once he or she has a partner, on he other hand, is
exercising individual agency.  In a choice of a safer sex practice such
as using a condom in a given situation, agency may be associated with
the pair of individuals and their negotiations.  

A natural agent such as a virus has its own sampling design.  In the
case of HIV it is a link-tracing design, reaching new individuals
through sexual and blood-exposure links.  The intervention is
infection of the individual.  

One intervention strategy interacts with another.  An intervention
design that follows links from individuals who test positive for HIV
is influenced in where it goes by the pattern of virus spread.  An
intervention such as an antiretroviral prescription or, potentially, a
vaccine or a cure, affects the virus by reducing its spread.
Behavioral changes of individuals and pairs in cooperation 
affect the spread of the virus.    The virus affects the population by
increasing mortality rate of infected individuals and, as described
later in this paper, affecting the link pattern or temporal network
geometry of the population.   Additionally, two types of virus such as
HIV and herpes simplex virus 2 can catalyze the spread of each
other.  

The sampling designs are adaptive in that the probability of following
a link or adding a unit to the sample can depend on the values of the
originating or candidate nodes, on values associated with the links
between, and on surrounding network conditions.  In addition, a deeper
adaptivity changes design parameters as virus strains evolve and
people learn.  

Because of the multiple agents, the interactions between designs, the
adaptations to interventions, and effects of interventions unfolding
over time, the births and deaths of nodes and the formings and
dissolvings of links, the problem is inherently complicated and it is
helpful to find simplifications.  A number of the designs, human and
natural, have common features.  Some variables in the overall process
have stochastic stability properties given specific intervention
strategies.  In such cases we can view the summary effect of an
intervention as the change in its equilibrium distribution once the
intervention strategy is put into effect.

\subsection*{Background}

% link tracing designs and network models

Design-based inference methods using Markov Chain Monte Carlo sampling
from a conditional distribution given a sufficient sample statistic
are used in link-tracing designs in networks
\cite{thompson2006aws,targetedwalks2006,thompson2011adaptive}.  The
current approach to likelihood-based inference with link-tracing
designs most often assumes design ignorability in the sense of
\cite{rubin1976inference}.  The likelihood approach to inference for
link-tracing designs in networks is described
\cite{thompson2000model}, developed for exact Bayes inference with a
simple model/design combination in \cite{chow2003estimation}, and for
computational Bayes inference with adaptive web sampling with a
stochastic block network model in Kwanisai, M. (2005, Estimation in
Link-Tracing Designs with Subsampling.  Ph.D. thesis, The Pennsylvania
State University, University Park, PA, USA].  This approach was
developed farther and for more complex network models in
\cite{handcock2010modeling} and subsequent papers by those authors. In
\cite{handcock2014estimating} the Bayes inference deals with both a
network model and a network sampling design that is nonignorable.
Issues of inference in network epidemic models are additionally
discussed in \cite{welch2011statistical}.

\cite{salehi2010empirical} describe empirical likelihood based
confidence intervals for adaptive cluster sampling.
\cite{deville2006indirect} analyze weighting systems for estimation
in indirect sampling, a class of adaptive network designs.

Approaches combining design and model based methods include the random
walk asymptotics-based estimators used with respondent driven sampling
\cite{heckathorn1997respondent,heckathorn2002respondent},
\cite{salganik2004sampling}, \cite{volz2008probability}, and
\cite{targetedwalks2006}.  A different
approach is used in \cite{félix2004combining} and
\cite{félixmedina2006combining}.  A different approach still
combining design and model based methods is developed in
\cite{gile2011network}.  Current methods in respondent driven
sampling are assessed in \cite{gile2010respondent},
\cite{goel2010assessing} and \cite{goel2009respondent}.

 \cite{rubin1978bayesian}, \cite{rosenbaum1983central}, \cite{thompson2002}, and other
authors have looked into the combined effect of experimental
treatment-assignment designs and unit-selecting
sampling designs on inference. 
A sample of units is selected from a population by some sampling
design.  This sample obtains the experimental units.  An experimental
design is then used to assign the experimental units to treatments.
The experimental design enables inference regarding the effects of
treatments to the experimental units, and the sampling design enables
inferring how this potential effect extends to the population as a
whole.

Recently a number of papers have come out with approaches to extending
static network models to the dynamic situation.
\cite{snijders1996stochastic,snijders2001statistical,snijders2002markov,snijders2005models}
and \cite{snijders2007modeling}
developed an interesting class of network evolution models based on
behavioral characteristics of actors/nodes such as tendencies toward
reciprocity, transitivity, homophily, and assortative matching.  At
the same time he developed inference methods to estimate model
parameters from incomplete longitudinal data.  A summary of this
work together with a review of other approaches to dynamic network
modeling is contained in \cite{snijders2010introduction}.
\cite{krivitsky2010separable} present a dynamic network model based on
exponential random graph models.  A recent summary of statistical
network models is provided by \cite{goldenberg2009survey}.

Latent space network models were introduced in 
\cite{hoff2002} for static networks. A dynamic extension of that
approach in which nodes  move by small random increments was developed
by \cite{Sarkar:2005:DSN:1117454.1117459}.  

In this paper I use the latent space approach with an underlying
temporal spatial point processes having grouping and clustering
tendencies and births and deaths of point-objects, which serve as the
network nodes.  The temporal network model on top of that has links
forming and dissolving over time, allowing for link probabilities to
depend on social distances between nodes, node characteristics
including including sex of a potential partner, group identities, and
various network properties such as degrees and component memberships
of each potential partner.  This model is then used for the spread and
adaptations of virus strains and evaluations of effectiveness of
different sampling designs and intervention strategies.

% network models and dynamic ones

%flame ranks
 %brin and page
 %russian authors recently on that

%ecf
 %epps

%ergodicity, stationarity of stochastic processes
 %parzen
 %a guy's book chapter i was using not long ago
 
%chimpanzees and siv relevant to equilibrium distribution

%rectangle tree, some phone thing

%left over citations:

% You may title this section "Methods" or "Models". 
% "Models" is not a valid title for PLoS ONE authors. However, PLoS ONE
% authors may use "Analysis" 
\section*{Methods}

In this project we are primarily interested in evaluating the
effectiveness of sampling designs for bringing interventions to units
in a population.  The intervention is then applied to units in the
sample, according to an intervention design.  From this point of view,
an experiment set in a population is an intervention with the purpose
of inferring a cause and effect relationship between the intervention
and its effect on sample units.  For an experiment it is considered
desirable that an intervention or treatment applied to one unit does
not affect the response of any other unit.  More broadly though, we are
particularly interested in interventions that, when applied to a
sample of units, have a desirable effect on the population as a whole,
including units to which the intervention has not been applied.  

The methods of the study consist largely of devising sampling designs
and intervention strategies for implementation in spatial, temporal,
and network settings.  An intervention strategy consists of a sampling
design for selecting the units to receive the intervention and a
procedure for assigning interventions to units in the sample.
Application of an intervention to a unit changes the value of some
variable associated with the unit.

We think of the population as a network or graph structure $G$
consisting of nodes and edges and having values $y$ associated with
nodes and $x$ associated with edges.   The network and its values
are stochastic and change over time, so that we have a stochastic process
\[
\{G_t, t \in T\}
\]
where $T$ is the time index set, which we will take to be discrete.
In more detail the population at time $t$ is 
\[
G_t = \{U_t, E_t, y(U_t), x(E_t)\}
\]
where $U_t$ is the set of units or nodes, $E_t$ is the set of edges or links
between nodes, $y(U_t)$ represents the variables of interest
associated with nodes, and $x(E_t)$ represents variables associated with
the links.  Note that $U_t$ and $E_t$ are random sets, containing
different elements at different time point, with insertions and
deletions of nodes and insertions and deletions of links taking place
over time.  

To make interventions in this population we need a sampling design to
reach nodes or edges on which to make the interventions.  At time $t$
the sample is a subset of the network and its values:
\[
s_t \subset \{U_t, E_t\}
\]
and we have a sampling process through time:
\[
\{s_t, t \in T\}
\]
The sampling design is the probability of selecting the sample, which
at time $t$ typically depends on the sample, the network and values at
times up to $t$, 
\[
{\rm P}(s_t \,|\, s_{t'}, G_{t'}, \phi_{t'}, t' \le t)
\]
where $ \phi_{t}$ are design parameters, which also may change with
time.   Once the sample at time $t$ is selected the intervention can
change values $y$ and $x$ associated with nodes or links in the sample.  
Note that over time the sampling process can select units to add to
the sample and units to drop from the sample.  

An intervention to a sample unit is a procedure that changes a value
$y_u$ associated with a unit $u \in s$, or a value $x_{uv}$ associated
with an edge $(u,v) \in s$ to a new value $y'_u$ or $x'_u$.  For
example if the intervention is prescribing a medication to person $u$,
an indicator variable $y_u$ might change from 0 to 1.  The
intervention is thus distinguished from the effects of an intervention
program, which develop over time and might affect nodes and edges outside
the sample as well as inside.   

The set of nodes $U_t$ is a stochastic point process.  We model it as
a spatial temporal point process with variable amounts of clustering
in space and motion over time.  
Specifically, we construct a spatial temporal Poisson cluster process
as follows.  A set of $k$ group center locations are initially selected
independently and uniformly over a study region, which we take as the
unit square.  
Let $c_{t}$ be center location at time $t$ for group a given group.  At each
time step $t$ it is perturbed as follows.  The most recent
displacement was $\delta_{t-1} = c_{t-1}- c_{t-2}$.  The new
displacement is   $\delta_{t} =\delta_{t-1} + \epsilon_t$.  Assuming
the spatial study region of interest is two dimensional, the random
perturbation $\epsilon_t$ is bivariate normal with means 0 a small standard
deviation equal in  each direction and zero correlation.  For a
process with directional motion  one can
add to the means or have unequal standard deviations and nonzero correlation.
This produces clusters that drift around independently and have a
momentum tendency in their ever changing direction and speed.
Further, this drift is kept in its initial distribution by using a
Markov Chain Monte Carlo (Metropolis-Hastings) selection step for each
group at each time step.  

The numbers of nodes per group are initially distribution
Poisson($\lambda$).  The expected number $\lambda$ can be constant or
selected independently for each group from a lognormal distribution,
producing more uneven sized clusters.  

The locations of nodes are initially distributed with a bivariate
normal distribution around group center locations.  Relative to the
group center, a node gets at each time step a random displacement
according to an autoregressive process or order 2, with the parameters
chosen to keep it in the initial bivariate normal distribution
relative to its group center.  Given the group centers, nodes move
independently to each other.  

Nodes are stochastically deleted over time depending on birth and
immigration rates and deleted depending on mortality and emigration
rates.  Mortality rates depend on things like ages and stages of
nodes.  Insertion rates are deliberately set to keep
the population and the group sizes in fluctuating but stochastically
stable distributions.  There are many ways to do this, as is done in
population dynamics models.  Here we simply imagine that as nodes die
or move out of the population, there is a tendency for new nodes to
move in over time to maintain a relatively stable population, so that
expected insertions is set accordingly.  Actual insertions are Poisson
with that mean value, and assignments of new nodes to groups is with
probabilities proportional to target group sizes compared to current
sizes.

Network links are inserted and deleted by the following process.  An
individual node has a selection function, centered on it and
decreasing with distance, for making tentative selections of nearby
nodes with whom to form a partnership.  Probability of selection
decreases with distance.  A simple choice is a normal kernel
function 
\[
g(d) = a_i e^{-d^2/2 \sigma_i^2}
\]
where $g$ is the probability of tentative selection by node $i$ of a
node $j$ at distance $d$ from it and $\sigma$ is a spread parameter which
can be node specific.  Further we set $g(d) = 0$ for some $d$ greater
than some maximum reach
distance such as $3 \sigma_i$.   Other selection function options
include the logistic function, which has an additional shape parameter
and the disk step function with is a constant out to the reach radius
and 0 beyond it.  

Meanwhile node $j$ may be making a parallel selection decision on node
$i$ and a link tentatively forms with probability 
\[
h(d) = a_ia_j e^{-d^2/2 (1/\sigma_i^2 + 1/\sigma_j^2)}
\]
In addition, the selection probability is further modified by
dependence on values of the two nodes, including the current degree
already of each, and by network values such as sizes of the components
each node is in and whether they are already in the same component or
not.  Dependence on a value such as degree can be compensatory, as
when a node will not take on a new partner if either it or the
potential partner has already one or more partners already, or
preferential attachment, when a node with high degree makes it more
likely to take on another relationship.  

In this way, the network formation process depends on $N_t$ designs,
where $N_t$ is the number of nodes at time $t$ and links are formed by
consensus or negotiations between pairs of nodes.  Deletion of a link
is at the discretion of either node in a partnership, and the
probability of deletion can similarly depend on current values.  

In this way the design that creates, maintains, and changes the
network is decentralized, with agency largely residing with
individuals and pairs.

The sampling designs with which we reach into the network population
can in some cases use a conventional frame such as a list of nodes and
in other cases use spatial sampling techniques.  A Bernoulli 
sample in a population of $N_t$ units has $N_t$ independent trials,
selecting unit $i$ with probability $p_i$, for $i = 1, ..., N_t$.  The
sample size is random, with expected value $\sum_{i=1}^{N_t}p_i$.  A
conventional design for sampling with replacement from $N_t$ units has
$n$ independent trials with unit $i$ having probability of selection
$p_i$ on each of the trials.  A random sample without replacement has
equal probability of selection for each possible combination of $n$
distinct units, giving probability $n/N_t$ probability that unit $i$
is included in the sample.  

Our focus of interest
in this paper is on the link-tracing designs.  These can be started
with any of the simple designs above.  In many cases of interest, the
sampling process attains over time a stochastically stable
distribution regardless of the initial design by which it started.
Also nodes are being

Consider a sampling process that at time $t-1$ has a sample $S_{t-1}$.
The sample $S_t$  is the result of following links out from $S_{t-1}$,
supplemented by new nodes selected at random, spatially, or other
design not relying on links.  An edge going from node $i$ to node $j$
is a pair $(i,j)$ in the current edge set $E_t$, the pair being
ordered if the link is directional.   

A flexible type of link tracing design
selects a Bernoulli sample of links out from the current sample to add new
nodes to the sample.  The probability $p_{(i,j)}$ of following link
$(i,j)$ to add node $j$ to the sample, for $(i,j)\in E_{t-1}$, 
$i \in S_{t-1}$, and $j \notin S_{t-1}$ can depend on values associated with
the origin node $i$, values associated with the destination node $j$,
and values associated with the link $(i,j)$ between them.  For a unit
$i$ outside the sample, the probability it is added at time $t$
through link tracing is 
\[
p(i \in s_t \,|\, i \notin s_{t-1}) = 1 - \prod_{\{j:j\in s_{t-1}, e(j,i) \in E_{t-1}\}} (1-p_{(i,j)})
\]
In addition
nodes may be added on occasion through direct Bernoulli sampling or
random sampling without replacement.

Nodes are removed from the sample through Bernoulli removals.  Thus
\[
p(i \notin s_t \,|\, i \in s_{t-1}) = q_i
\]
The probability $q_i$ for removing node $i$ from the sample may depend
on values associated with the node such as how long it has been in the
sample.  In many cases the 
Bernoulli removals may be independent from one sample node to another.  In other cases, however,
dependence results from limited resources or designs which increase or
decrease probabilities of removal depending on current sample size. In addition a node $i$ is removed from the sample when it is
deleted from the population.  

Different variations of this type of design can select a random sample of
without replacement of $n_t$ links out from the sample, where $n_t$
may have a fixed target value but be constrained to be no greater than
the number of links out. If the number of links out is less than
$n_t$, the additional desired units can be selected by simple random
sampling from those not in the sample.   In another variation links out are selected
with replacement using $n_t$ independent trials with link  $(i,j)$
having probability of selection $p_{(i,j)}$
for $(i,j)\in E_{t-1}$, 
$i \in S_{t-1}$, and $j \notin S_{t-1}$.  Further variations include
designs of a type used in respondent driven sampling of members of a
hidden population in which each person in the sample is given a set
number $k$ of recruitment coupons with which she can recruit up
to $k$ individuals with whom she is linked.  

In the temporal setting, sampling without replacement has more than
one potential meaning.  It can mean that we do not select a unit that
is already in the current sample, and keeping track of the number of
times a unit is selected or else only keeping track of the set of
distinct units selected.  Alternatively, without-replacement sampling
can mean, after a unit has been removed from the sample, we will not
later select it back into the sample.  Further we can generalize the
concept of replacement to a continuous variable ranging from 0
to 1.  In that case we multiple the usual selection probability $p_i$
for unit i times $r$, so that $r=1$ corresponds to complete
replacement and $r=0$ is strictly without-replacement.  An
intermediate value of $r$ means that selection probability is damped
for a unit previously in the sample.  In addition $r$ can depend on
how long it has been since the unit was last in the sample.

A random walk in a graph is a simple design that has received much
study in the static network setting but little in the temporally
stochatic graph setting.  
A simple random walk design has one unit in the sample at a time.  One
link is followed at random to find the new sample unit, and the
current one is dropped.  Letting $d_{ti}$ be the degree, or number of
links out from node  $i$ at time $t$, 

\[
p(s_t = \{i\} \,|\, s_{t-1} = \{j\}) = 1/d_{ti}
\]

This is often supplemented with a probability $p$ of taking a random
jump to a different unit in the population, giving 
\[
p(s_t = \{i\} \,|\, s_{t-1} = \{j\}) = p(1/d_{ti}) + (1-p)(1/(N_t-1)))
\]
so that the walk does not get stuck in a single component of the
network.  The classical random walk is with replacement, so that in
the static graph setting it forms a Markov chain with the current
state of the process being the node in the current sample.  This
viewpoint runs into some complications in the temporal stochastic
graph setting of interest here.  The nodes themselves are transient
entities, because of the birth and death process, so that they can not
form recurrent states of a stochastic process.  Further, we can follow
a link to a node, only to have the link be deleted at that time step,
so we may have the sample stuck for a long time on a single isolated
node until a new link might connect to it.  Thus the design might be
modified to take a random jump with some higher probability if that
happens.  Still further, the current node may itself be deleted while
it is the sample node, and the sampling process will have ended unless
we again modify the design to start a new random walk from a randomly
selected node.  Also, as network links change the connected components
of the graph change, and a simple random walk that is stuck in one
component can move to a different component when they transiently
merge.   

The population process together with the sampling process gives us a
complex stochastic process.  Even though what would appear to be basic
entities of the process, like nodes, links, and sample members are
transient, we find that under a wide range of conditions stochastic
stability properties of some variables are apparent.  By stochastic
stability we refer to variables that have stationary distributions or
ergodic properties over time.  A variable in such a distribution is
forever varying, or fluctuating, but it's distribution stays the same,
once equilibrium is reach.  

Although it is not necessary to have stationary of limiting
distribution of variables in order to study the effectiveness of
intervention strategies, it is convenient in a number of ways.  With
stochastic stability, population values will stay in a predictable
distribution in the absence of any interventions.  If we make an
intervention, a simple measure of its effect is the equilibrium
distribution that results over time, in comparison with the
equilibrium distribution without the intervention.

In many cases an agent we make an intervention against, adapts over
time to counter our intervention.  More generally, one intervention
design interacts with another.  The effect of the intervention is
most simply measured as the equilibrium that results after all
adaptations and interactions attain their new equilibrium
distributions.  

In more detail, the effect of an intervention is the distribution of
sample paths over time, compared to the distribution of sample paths
without the intervention, regardless of whether the processes are in
equilibrium.  In terms of a simulation study, an advantage of
processes that attain equilibrium distributions of effects over time
is that the distributions can be determined from a single realization
over a long time span, or a few such realizations, rather than in
every case needing to be run for a large number of realizations.

Because the overall stochastic process is complex and changes in
values of input parameters can change a process from stable to
non-stable, we examining stability properties empirically.  Helpful
tools for doing this include time series plots of variables of
interest, cumulative mean functions of such functions, cumulative
histograms, and cumulative empirical characteristic functions (ECF) of key
variables.  The empirical characteristic function of a stochastic
process variable $X_t$ is defined as 
\[
c_x(a) = \frac{1}{t}\sum_{t'=0}^te^{iaX_{t'}} =  \frac{1}{t}\sum_{t'=0}^t
\left[\cos(aX_{t'}) + i \sin(aX_{t'})\right]
\]
We are interested not only in when the process is stochastically
stable, but when it is changing.  In particular, once we initiate an
intervention strategy, we would like to see early signs that it is
changing.  The tail of the ECF is sometimes described as reflecting
the roughness of a distribution.  In many cases when a process starts
to change, as the distribution starts to change in the direction of
the new equilibrium, the tail of the ECF appears to go wild, writhing
like a snake that has become restless, calling attention to the
change.

Many parameters of the network model and designs are individual, by 
node.  They can be changed for one individual in the course of a
simulation run.  In this case we can ask, what is the benefit to that
a single individual of making this change.  Individual parameters
include tendencies in forming links, compensatory changes for high
degrees of self or other, or preferential attachment tendencies.  

In designing interventions against a virus spreading in a human
network, we would like to be able to anticipate or infer where the
virus might spread next, and which units to distribute our
intervention to in order to have the most beneficial effect.  To do
this we devise a simple type of design based inference in the
network.  This is done as follows.  

Consider the link tracing design that was described above that samples
with replacement regarding units that were previously in the sample
and without replacement regarding units currently in the sample.
Bernoulli tracing of links, independently with probabillity $p$ for
each link leading out of the sample, is supplemented by some small
chance for selecting random units, which is done with independent
Bernoulli selections from the units outside the sample with small
probability $p_r$.  Given the current sample size $n_t$ after link
tracing additions and a target sample size $n_{\rm target}$, units are
removed from the sample by independent Bernoulli removals, each having
probability $q$, where $q = (n_t - n_{\rm target})/n_t$ if $n_t -
n_{\rm target} > 0$ and $q = 0$ if $n_t - n_{\rm target} \le 0$.  If
we give the design a high rate $p$ of tracing links out, the sample
moves very fast through the population.  Its sample size varies but
stays in a stable distribution around the target size.

Hazard function based on Bernoulli (p) per time step has expected time
to event $E(x)=1/p$. 
Hazard function based on a Weibull distribution discrete approximation,
\[
f(x) = \lambda\beta \lambda^{(\beta - 1)} e^{(-x\lambda)^\beta}
\]
\[
     E(x) = \Gamma(1 + 1/\beta)/\lambda
\]
 set $E(x) = 1/p$ and solve for $\lambda$, giving

\[
\lambda = p \, \Gamma(1.0 + 1.0/\beta)
\]
\[
     h(x) = (\lambda \beta)  (\lambda  x)^{(\beta - 1)}
\]

A daily probability $p$ of an event such as mortality based on a
longer term rate $p_a$ such as probability per year is calculated from 
$
p_a = 1 - (1-p)^k
$
or its inverse 
\[
p = 1 - (1-p_a)^{1/k}
\] 
where $k$ is the number of time steps in the longer time rate, such as
365 days in a year.

\subsubsection*{HIV Epidemic}

The HIV epidemic serves as a motivating example for the methods of
this project.  We focus in particular on the heterosexual epidemic.
The virus uses a link-tracing design to select a sample of people.  It
makes an intervention by infecting each person in its sample.  The
links followed are sexual partnership links, and transmissions provide
the link tracings.  The probability of tracing is very low per contact
event, often but not always less than one one-thousandth.  We model
the design as without replacement, though that is a simplification
since in some cased there may be multiple infection of an individual
with different strains of virus.  Since there is currently no
practical cure available for HIV, the design is without replacement.
Removal of a node from the sample occurs only with deletion of the
node from the population, at death or emigration out of the study
population.  The probability of tracing depends on values of the
origin node, the destination node, and the link between them.  For
example, the probability of transmission of HIV per sexual contact
event depends on the stage of infection in the infected individual,
the susceptibility to infection of the exposed individual, and on the
details of the type of contact in that event.  

For the virus' design parameters, there is a tradeoff in which a high
tracing rate is associated with a decrease in survival time of its
host.  Letting $\alpha$ represent virulence or host mortality rate and
$\beta$ represent transmission rate, a simple form of function
characterizing the tradeoff is (Boyker, Fraser)
$\beta = c \alpha ^{1/\gamma}$
where $\gamma > 1$.  Thinking of virus modifying its transmission rate
by small increments over time, with the per-time-step mortality rate
of the host human being affected as a result, the relationship can be
written 
\[
\alpha_t = a \beta_t^{\gamma}
\]
where $a$ is a constant
related to $c$ and $\gamma$ is a curvature parameter.

Existing and potential
interventions to reduce the epidemic include behavior changes
affecting patterns of relationships between individuals, safer sexual
practices used strategically, antiretroviral drug combination
treatments, vaccines, cures, pre-exposure prophylactic treatments for
partners of infected individuals, and control of catalyzing infections
such as Herpes simplex virus 2 (HSV2).  In counter-response, HIV
adapts to interventions against it with mutations, recombinations, and
their selections.      

\subsubsection*{Network epidemic dynamics and rates background}

% hiv transmission rates

Rates of heterosexual transmission of HIV per coital act and frequency of coital acts are
estimated in \cite{gray2001probability} in a study of 171 monogamous
couples in which one member was HIV positive, in Rakai, Uganda.
Probability of transmission to
the other partner in the longitudinal study was estimated as a function
of viral load.  Rates were about 4 times higher is the presence of
genital ulceration disease (GUD).
\cite{hollingsworth2008hiv} have estimates for stage specific rates of
HIV transmission based on serodiscordant heterosexual couples in Rakai,
Uganda.  They estimate a 26 times for early stage and 7 times for late
stage compared to chronic stage.  The estimate early stage infection
lasts about three months.  
For this study I've used initial estimates of HIV transmission rates with and without
catalyzing genital ulceration disease (GUD) such as herpes simplex
virus type 2  
(HSV-2) from the metastudy \cite{boily2009heterosexual}.

Effectiveness of condom use in reducing heterosexual HIV transmission
is reviewed in \cite{welleret2007}.  

% early rate

Infectious agents typically run into a tradeoff in the evolution of
transmission rates.  Higher transmission rate is associated with
increased virulence, increasing the mortality rate of the host which
in turn slows down the spread of the pathogen strain in the
population.  Tradeoff functions have the general form
that transmission rate is a decelerating function of
virulence. 
In this study I use a simple tradeoff function from
\cite{bolker2010transient}, who uses rate data from
  \cite{fraser2007variation} for HIV.  Other functions with similar tradeoff
  properties have been used by other authors such as
  \cite{nowak1994superinfection}.  

 \cite{cohenet2012} presents
opposing views of different researchers on the relative importance of
early stage in the transmission of HIV.     \cite{cohenet2012} presents
opposing views of different researchers on the relative importance of
early stage in the transmission of HIV.    

Dynamic epidemic models
showing selective advantage of virulent strains in early stage of an
epidemic and advantage shifting to less virulent strains, which allow
their host to survive longer, as the epidemic matures were compared to
laboratory studies with colonies two competing strains of bacteria
\cite{Berngruber2013}, finding agreement with the model predictions.
 \cite{cohenet2012} presents
opposing views of different researchers on the relative importance of
early stage in the transmission of HIV.

\cite{lewin2011finding} describes some of the challenges in developing
an HIV cure based on antiretrovirals together with an agent for
releasing the virus from latency.  

%dynamic epidemic models

 Properties of networks in which
relationships shift preferentially that are missed by static network
epidemic models are examined in \cite{fefferman2007disease}.  The
stability of casual contact and close contact patterns over time was
studied \cite{read2008dynamic} using diary based methods with 49
volunteers, finding that the close contacts tended to be more stable
than casual contacts. 

Network structure and change patterns in relation to individual
behaviors were investigated in \cite{potterat1999network} for
syphilis among young people and HIV among drug users, finding in both
studies that spread of disease was associated with network cohesion,
in the form of separation of components or local density of
connections.

The importance of concurrent relationships in the spread of the HIV
epidemic is investigated in \cite{watts1992influence} and
\cite{morris1997concurrent}.  Interaction of early transmission rate
and concurrency in sexual links is discussed in
\cite{eaton2011concurrent}.  A modeling approach combining network
models and compartment models is used in \cite{goodreau2010concurrent}
to evaluate the importance of concurrency.

\cite{kamp2010} describes a modeling approach based on a set of
partial differential equations with the addition of some temporal network
aspects such as degree distributions in which contacts change,
describing mean behavior for infinite network size and approximate
behavior for moderate size. 
\cite{rocha2011simulated} and \cite{rocha2012epidemics} describe a
network of 50,185 sexual contacts between 6,642 escorts and 10,106 sex 
buyers as reported on a Web discussion forum.

%adherence and its pattern - maybe change to abc 

\cite{phillips2013increased} (and especially its Supplement 1) modeled the individual variability of
adherence over time for an individual as well as the variability
between individuals.  Their model, used for the purpose of estimating
parameters, is individual based but not network based except to the
extent of having assumed degree distributions.  This study uses an
inference method that appears to essential be approximate Bayesian
computation.  

An approach bringing network effects to epidemic compartment models by
having different groups with different network degrees is described in
\cite{house2011insights}.  Epidemic threshold properties of simple dynamic
network models in which degree says constant but neighbors exchange,
that is, identity of partners change instantaneously at random times
are described in \cite{volz2009epidemic}.  Compartmental models are
modified to add some network effects in \cite{bansal2007individual}.
  \cite{gupta1989networks} examine network effects in
compartmental models by including a contact or mixing matrix into the
model, comparing assortative mixing patterns, in which individual's
contacts tend to be within their own group and dissortative patterns,
in which contacts tend to be between groups.  Early work in modeling
the dynamics of the HIV epidemic includes 
\cite{may1987transmission, may1988transmission} and 
\cite{anderson1986preliminary}.   

The models for dynamic spatial network populations and interacting
designs developed in work also have some relationship to the
literature on evolutionary dynamics.  \cite{nowak2006} provides a summary
of models of species interactions, epidemics, and selection based on
systems of partial differential equations.  \cite{champagnat2006unifying}
describe recent work on stochastic point process models built on top
of that approach. Among the many cases of ecological systems
exhibiting the sort of spatial, temporal, and network patchiness
addressed by the models and designs of this paper, a good example
is described by \cite{rudicell2010impact}.

%\subsubsection*{Computing note}

% Results and Discussion can be combined.
\section*{Results}

In this section we look at some results of using the network sampling
approach to evaluate the effectiveness of intervention designs, using
the HIV epidemic as our test example.  The first type of result we
look at are what are the characteristics of the virus design and how
does it respond to human network activity over time.  The second
result we look at is what can one individual or two in cooperation
accomplish to reduce the risk to themselves or others.  And the third
result we look at involves network effects of two types of ideal
interventions we hope are available at a future date, namely
treatments that cure or clear an infected individual's body of HIV.  

\subsubsection*{Virus adaptation to temporal
  network geometry}

Most link-tracing designs through networks select nodes having more
links in to them with higher probability than nodes with fewer links.
This is true of simple random walk designs, snowball designs of
various types, and adaptive web sample designs among others.  It is
true of a natural link-tracing design such as HIV uses in selecting
people following sexual links, and which we model here as independent
Bernoulli selections, without-replacement, with unequal probabilities
of transmission tracing depending on node and like characteristics.
The only designs I am aware of for which this is not true are targeted
random walk designs in which a random walk in a graph is modified
using Markov chain Monte Carlo techniques in order to achieve desired
long term selection probabilities including having them equal for all
nodes.  

Combined with with network clustering tendencies in which certain sets
of nodes are at least temporarily more highly connected than average,
the link-tracing design of the virus leads to a pattern of spread in
the population over time characterized by local explosions, even if
the explosions occur in relative slow motion, followed by longer
periods of little spread.  We can document this trend by tabulating
the average degree of nodes recently selected into the virus' sample,
compared to the average degree of nodes not in the sample and to nodes
that have been in the sample for longer.  We find the degree of nodes
recently selected is higher.  The same is true for their out-degree,
the number of links a node in the sample has to nodes not in the
sample.

In a node that has recently been infected, a strain of virus with a
high transmission rate will have an advantage, on average, relative to a strain
with a low rate.  The high rate strain can take advantage of the high
number of links out to spread farther.  There is a tradeoff, however,
since the higher virulence associated with the high transmission rate
brings a higher mortality rate to the host person.  Later in the same
person, there tends to be fewer if any links out to nodes not already
infected.  At that stage a strain of virus with low transmission rate
is favored.  With the longer expected survival time of the host, the
virus has a chance over time of new links being formed.   
 
The advantage to the virus of having a high transmission rate early
for a short time and a low transmission rate later for a long time is
amplified by clustering of network links in time and space.  As
infection spreads through a cluster a strain with higher transmission
rate, particularly higher early rate, will spread explosively through
the cluster, faster than other strains, and become locally more
predominant.  Over time their are fewer links to uninfected nodes in
the cluster, as more of the nodes locally are infected.  Over a longer
period of time the cluster breaks up through social drift.  Survival
time of a host person is higher for a low virulence, which gives that
strain a better chance of being the lead strain in igniting the next
cluster, should it be encountered in the drift over time as old links
dissolve and new ones are formed.

The pattern of simulation results comparing strains adds further light
on this issue.  Comparing strains with fixed early and chronic rates,
an optimal strain having a transmission probability per sexual contact of
about 0.008  during chronic stage and about 15 times that during early
stage emerges.  Compared to a strain with the same chronic rate and an
early rate the same as that, that is, an early rate factor of 1, the
strain with low early rate reaches over time reaches an equilibrium
presence about ten percent lower.  More noticeably, the low early rate
strain tends to take much longer to get off the ground, starting from
just one or a few cases.  A strain with a very high early rate of 50
times or more tends to rise very fast and then crash to a lower
equilibrium because of the higher host mortality induced.  Strains
with an early factor between about 5 and 35 perform about as well as
the optimal strain, in terms of equilibrium level.  In each run a
single strain exists in a single population.

When we put two strains with fixed rates together in competition, the
optimal strain almost always nearly completely dominate over time over
either the one with low early rate or very high early rate.

Next we put in a random selection of strains and let them evolve
together.  When a strain transmits to a new node, the new early
transmission rate is the value of the one in the transmitting node
plus a small random increment (uniform or normal), constrained to not
go below zero or above 1.0.  The result, over time, is independent of
the initial distribution and tends to produce a variable distribution
of early rates in the population, with mean early rate factor between
15 and 20 and the bulk of the distribution between 5 and 40.
Contributing factors to the persistence of variability of early rate
in the population are the variability at transmission, producing
genetic drift, and the fact that at any given moment different nodes
are under different selective pressures, depending on the number of
open links around them.  

The two-pronged strategy of HIV with a high early transmission rate
for a short period followed by a low transmission, low mortality rate
for a longer period, presents challenges for interventions to lower
the incidence and prevalence of HIV in a human population.  The
highest transmission rate comes before the virus can be detected with
standard tests.  Explosions into concentrated clustered tend to be
well established before responses can be readied.  And in the survival
period that that averages around ten years without treatment only a
few of those cases need to make an ignition of a new explosive area
for the epidemic to persist.

\subsubsection*{Pair consensus intervention strategy of safer sex early in
relationship followed by HIV home testing}

Individuals and pairs of individuals have the most direct agency in
forming and dissolving links, which creates and maintains the network
over time and potentially can change it.  As described earlier we
think of this as $N_t$ designs, as many as there are individuals.  The
seemingly insignificant day to day choices and actions of individuals
create the uneven temporal network topology in which the HIV epidemic
expands or shrinks locally and throughout the world.  

In this section we look at a strategy that relies on the agency of
individuals and cooperating pairs.  The idea is to counteract the
virus strategy with it's high rate of transmission in the early stage
of an infection.  The problem is, the standard tests of for the
presence of HIV are not sensitive in the early part of an infection,
and the person having an infection in early stage is likely not to
know they are infected.  Instead, we consider the strategy  in which a
pair uses safer sex practices early in a relationship, followed by
each person testing the other.  We consider the use of one of the new,
cheek swab home tests.  

The strategy is as follows.  The couple use safer sex practices during
the first k weeks of a relationship, followed by HIV tests.  If both
tests are negative, then the restriction to safer sex practices is
lifted.  If both tests are positive, then also the pair can abandon
the safer sex practices with each other.  If one test is positive and
the other negative, the pair continues to use the safer sex
practices.  

A safer sex practice is one that reduces probability of transmission
in a contact to a specified proportion $p_s$ of penetrative
intercourse.  In the simulations on the heterosexual epidemic I
use the value $p_s = 0.10$.  The literature is sparse but 
appears that not just the use of a condom but alternatively a range of
sexual techniques collectively referred to as ``outercourse'' reduce
transmission probability to something like 10 percent of what it would
be.  These include various methods of oral sex and hand-genital
contacts.  We omit from the list fellatio involving ejaculation into a
partner's mouth, as one study estimated is transmission rate as 50
percent that of penetrative sex.  

For the duration of the safer sex period from the start of a
relationship the simulations use 12 weeks, or 84 days.  The expected
duration of early stage infection from the start of HIV infection in
the simulations is 75 days, or between 10 and 11 weeks.  In this way
the infection has a high probability of being detectable when the
tests are given and the chance of being exposed to the high
transmission rate from a new partner is greatly reduced.  

In simulations when everyone in a population is using this strategy it
is surprisingly effective in bringing down the epidemic even in dense
preferential attachment situations.  Suppose only 50 percent, or 10
percent of the population uses this strategy.  What is the effect on
them, what is the effect for others.  The answer depends on the
pattern of adoption of the strategy.  If individuals with the strategy
tend to assort to relationships with others using the same strategy
the benefit for them is great and others are relatively unaffected.
If the two types of behaviors mix randomly to benefit to the people
using it is dampened and others accrue benefit.  We can this type of
exploration to the extreme and ask, what is the effect on one person
if she uses this strategy while others do what they will do anyway,
including dense, high risk social settings.  To use this strategy
takes the cooperation of both partners in a relationship.  If her
partner has other relationships also, her risk is affected if the
partner uses the strategy with just her or with their other partners
as well, at least while they are in a relationship with her.  

The concept of a strategy of this type is that individuals and
cooperating pairs bring down the potential rates of transmission in
relationships, with particular focus on early rates and untestable
periods.

\subsubsection*{Two types of cures, one conferring immunity}  

At the time of writing there is no practical cure for HIV infection,
that is, no treatment that will clear the virus completely from the
body and leave the person in good health.  There is, however,
plausible theory on approaches for developing cures for HIV.  One
person has apparently been cleared of the virus through stem cell
transfusion giving him immune cells having a protective mutation but
for various reasons it is not considered practical to spread this
approach widely.  A cure would be the the best thing that could happen
to an individual who is infected with HIV and in our simulations a
cure emerges as having the best network dynamics for bringing the
epidemic down through prevention of further spread, even in very
difficult network situations. 

In the simulation two types of cures, both equally effective in curing
an individual, emerge as having markedly different network dynamics.
One type of cure clears the virus from the person but leaves him or
her susceptible to reinfection.  The other type of cure, in addition
to clearing the virus, confers subsequent immunity to infection.  In
between are cures that would confer some degree of immunity, which is
incorporated as a resistance factor between 0 and 1.  

One research
approach uses antiretroviral drugs to deeply suppress reproduction of
HIV particles in the blood or other body fluids and seeks another
class of drugs to induce the virus in refuge as protovirus segments of
human DNA to express themselves and emerge to be in turn prevented
from further reproduction.  A second research approach in seeking a
cure involves gene therapies to insert strengthening proteins or other
HIV resistant features in immune cells.  There is no apparent reason
to anticipate that the first type of cure would confer immunity
whereas the second type could reasonably expected to confer at least some
degree of immunity following cure.   

Each of the interventions studied in the simulation can be effective
in bringing down the epidemic in many network settings and can be
overwhelmed by others.  The most difficult of the temporal networks in
the simulation are dense and highly clustered in terms of links, and
are represented in the simulations by preferential attachment
tendencies in the formations of links, giving a high degree of
clustering in space and in time.  What happens with the cure that does
not confer immunity is that the high-degree, highly connected, or
high-change individual who is most likely to be infected and serve as a
conduit of transmission to others, once cured is likely to be
reinfected and again serve as a conduit as least over near time.  The
cure of an individual by the second type of cure not only reduces
prevalence in the population by one but also makes the links of that
person unavailable or unconducive for further spread.

%\section*{Discussion}

% Do NOT remove this, even if you are not including acknowledgments
\section*{Acknowledgments}
This work is supported by the Natural Science and
  Engineering Research Council of Canada.  

%\section*{References}
% The bibtex filename
\bibliography{refs2015}

%\section*{Figure Legends}
%\begin{figure}[!ht]
%\begin{center}
%%\includegraphics[width=4in]{figure_name.2.eps}
%\end{center}
%\caption{
%{\bf Bold the first sentence.}  Rest of figure 2  caption.  Caption 
%should be left justified, as specified by the options to the caption 
%package.
%}
%\label{Figure_label}
%\end{figure}

%\section*{Tables}
%\begin{table}[!ht]
%\caption{
%\bf{Table title}}
%\begin{tabular}{|c|c|c|}
%table information
%\end{tabular}
%\begin{flushleft}Table caption
%\end{flushleft}
%\label{tab:label}
% \end{table}

\end{document}